\documentclass[
12pt,
preprint,preprintnumbers,nofootinbib,
groupedaddress,superscriptaddress,amsmath,amssymb]{revtex4}
\usepackage{graphicx}
\usepackage{dcolumn}
\usepackage{bm}
\usepackage{amssymb}
\usepackage{amsmath}
\usepackage{epsfig}    
\usepackage{color}
\usepackage{hhline}

\def\be{\begin{equation}}
\def\ee{\end{equation}}
\newcommand{\bea}{\begin{eqnarray}}
\newcommand{\eea}{\end{eqnarray}}
\newcommand{\nn}{\nonumber}

\numberwithin{equation}{section}

\begin{document}
{\begin{flushright}{APCTP Pre2020 - 005}\end{flushright}}

\title{A radiative seesaw model with three Higgs doublets\\
 in modular $A_4$ symmetry}

\author{Hiroshi Okada}
\email{hiroshi.okada@apctp.org}
\affiliation{Asia Pacific Center for Theoretical Physics (APCTP) - Headquarters San 31, Hyoja-dong,
Nam-gu, Pohang 790-784, Korea}
\affiliation{Department of Physics, Pohang University of Science and Technology, Pohang 37673, Republic of Korea}

\author{Yutaro Shoji}
\email{yshoji@kmi.nagoya-u.ac.jp}
\affiliation{Kobayashi-Maskawa Institute for the Origin of Particles and the Universe, Nagoya University, Nagoya, Aichi 464-8602, Japan}

\date{\today}

\begin{abstract}
 We propose a radiative seesaw model based on a modular $A_4$ symmetry,
which has good predictability in the lepton sector.  We execute a
numerical analysis to search for parameters that satisfy the
experimental constraints such as those from neutrino oscillation data
and lepton flavor violations.  Then, we present several predictions in
our model that originate from the modular symmetry
{at a fixed point as well as fundamental region of $\tau$. }
\end{abstract}
\maketitle
\newpage

\section{Introduction}

One of big mysteries in the standard model (SM) is the origin of the
flavor structures. In particular, the flavor structure of the neutrino
mass matrix is very important to understand the lepton
sector. Historically, models with non-Abelian discrete flavor symmetries
have been widely discussed since they not only reproduce the
experimental results but also have several model specific predictions.

Recently, starting from papers~\cite{Feruglio:2017spp,
deAdelhartToorop:2011re}, modular motivated non-Abelian discrete flavor
symmetries have attracted attention of many authors to realize more
predictable flavor structures in the quark and the lepton sectors.  One
of their remarkable natures is that any coupling constants as well as
fields can also be transformed as non-trivial representations of
those symmetries.  Thus, we do not need to introduce many scalar fields
such as flavons to realize flavor structure. As a result, we obtain a
more minimal scenario without assumptions such as vacuum alignments
among scalar fields.

Here, we list several references where they apply this kind of
symmetries to flavor models;
the $A_4$ modular group~\cite{Feruglio:2017spp, Criado:2018thu, Kobayashi:2018scp, Okada:2018yrn, Nomura:2019jxj, Okada:2019uoy, deAnda:2018ecu, Novichkov:2018yse, Nomura:2019yft, Okada:2019mjf, Ding:2019zxk, Nomura:2019lnr, Kobayashi:2019xvz, Asaka:2019vev, Zhang:2019ngf, Gui-JunDing:2019wap, Nomura:2019xsb, Kobayashi:2019gtp, Wang:2019xbo, King:2020qaj, Abbas:2020qzc}, 
$S_3$~\cite{Kobayashi:2018vbk, Kobayashi:2018wkl, Kobayashi:2019rzp, Okada:2019xqk}, 
$S_4$~\cite{Penedo:2018nmg, Novichkov:2018ovf, Kobayashi:2019mna, King:2019vhv, Okada:2019lzv, Criado:2019tzk, Wang:2019ovr}, 
$A_5$~\cite{Novichkov:2018nkm, Ding:2019xna, Criado:2019tzk}, 
larger groups~\cite{Baur:2019kwi}, 
multiple modular symmetries~\cite{deMedeirosVarzielas:2019cyj}, and
double covering of $A_4$~\cite{Liu:2019khw}, in which masses, mixing, and
CP phases for quark and lepton are predicted~\footnote{Some reviews
\cite{Altarelli:2010gt, Ishimori:2010au, Ishimori:2012zz,
Hernandez:2012ra, King:2013eh, King:2014nza, King:2017guk,
Petcov:2017ggy} are useful for the understanding of non-Abelian
groups and their applications to flavor structures.}.
A possible correction from K\"ahler potential is also discussed in
Ref.~\cite{Chen:2019ewa}.  Furthermore, a systematic approach to
understand the origin of CP transformations is recently discussed
in ref.~\cite{Baur:2019iai}, and CP violation in models with modular
symmetry is discussed in Ref.~\cite{Kobayashi:2019uyt, Novichkov:2019sqv}.

Another big mystery in the SM is the lack of a dark matter (DM)
candidate.  Even though many experiments from different aspects are
going on to search for DM signatures, we have not obtained any decisive
proofs yet.  However, there are a lot of nice scenarios of DM that are
connected to other observables.  One of interesting models is known as
the radiative seesaw model~\cite{Ma:2006km}.  This scenario not only
explains the neutrino sector and DM at the same time but also provides a
lot of new phenomena at a low energy scale such as lepton flavor
violations, muon anomalous magnetic moment, collider signatures,
etc. Since such a model connects the DM sector and the neutrino sector,
the understanding of the neutrino nature leads to the understanding of
the DM nature, and vise versa.

In this paper, we work on a radiative seesaw scenario with a Dirac DM
candidate based on our previous work~\cite{Okada:2020oxh}, applying a
modular $A_4$ flavor symmetry. Then, we try to find several predictions
in the lepton sector. {Also we explore the parameter region
around a fixed point and show its prediction.  }
 
The manuscript is organized as follows.  {In Sec.~\ref{sec:realization},
we give our model setup under the $A_4$ modular symmetry, in which we
review the modular $A_4$ symmetry and define relevant interactions needed
to formulate the neutrino mass matrix, lepton flavor violations
(LFVs), and relic density of our DM candidate.  Then, we execute a numerical analysis and give several
predictions in the lepton sector in Sec.~III.  Finally, we give our
conclusion and discussion in Sec.~\ref{sec:conclusion}.}

\section{Model}\label{sec:realization}
In this section, we introduce our model, which is based on a modular
$A_4$ symmetry.  The leptonic fields and the scalar fields of the model,
their representations under the $A_4\times Z_3$ symmetry and their
modular weights are given in Tab.~\ref{tab:fields}. 
{The $Z_3$ symmetry is required to retain the Dirac nature of $N$
since modular $A_4$ symmetry cannot forbid Majorana terms, $\bar N_R N^C_R$ and  $\bar N_L N^C_L$.}
We also show the
representations of the Yukawa couplings in Tab.~\ref{tab:couplings}.
Under these symmetries, we write the renormalizable Lagrangian for the
lepton sector as follows:
\begin{align}
-{\cal L}_L &=
\sum_{\ell=e,\mu,\tau}y_\ell \bar L_{L_\ell} H_{SM} e_{R_\ell}\nn\\
&\hspace{3ex}+\alpha_\nu \bar L_{L_e} (Y^{(2)}_{\bf 3}\otimes N_{R})_{\bf1}\tilde H_1
+\beta_\nu \bar L_{L_\mu}(Y^{(2)}_{\bf 3}\otimes N_{R})_{\bf1''}\tilde H_1
+\gamma_\nu \bar L_{L_\tau}(Y^{(2)}_{\bf 3}\otimes N_{R})_{\bf1'}\tilde H_1\nn\\
&\hspace{3ex}+a_\nu (\bar N_{L_e} \otimes Y^{(6)*}_{\bf 3})_{\bf1} {L^C_{L_e}}\tilde H_2
+b_\nu (\bar N_{L_\mu}\otimes Y^{(6)*}_{\bf 3})_{\bf1'} {L^C_{L_\mu}}\tilde H_2
+c_\nu (\bar N_{L_\tau}\otimes Y^{(6)*}_{\bf 3})_{\bf1''} {L^C_{L_\tau}}\tilde H_2\nn\\
&\hspace{3ex}+a'_\nu (\bar N_{L_e}\otimes Y'^{(6)*}_{\bf 3})_{\bf1} {L^C_{L_e}}\tilde H_2
+b'_\nu (\bar N_{L_\mu}\otimes Y'^{(6)*}_{\bf 3})_{\bf1'} {L^C_{L_\mu}}\tilde H_2
+c'_\nu (\bar N_{L_\tau}\otimes Y'^{(6)*}_{\bf 3})_{\bf1''} {L^C_{L_\tau}}\tilde H_2\nn\\
&\hspace{3ex}+ {M_D} (\bar N_{L}\otimes N_{R})_{\bf1}
+ {\rm h.c.}, 
\label{eq:lag-lep}
\end{align}
where $\tilde H\equiv i\sigma_2 H^*$ with $\sigma_2$ being the second
Pauli matrix and $(A\otimes B)_{\bf R}$ indicates that the
representation, $\bf R$, is contracted from $A$ and $B$. Here, $M_D$
includes a modular invariant coefficient, $1/(i\tau-i\tau^*)$, and the
charged-lepton matrix is diagonal thanks to the $A_4$ symmetry.

\begin{center} 
\begin{table}[tb]
\begin{tabular}{||c||c|c|c||c|c|c||}\hline
&\multicolumn{3}{c||}{Fermions} & \multicolumn{3}{c||}{Bosons} \\ \hline \hline
  & ~$(\bar L_{L_e},\bar L_{L_\mu},\bar L_{L_\tau})$~ & ~$(e_{R_e},e_{R_{\mu}},e_{R_{\tau}})$ ~ & ~$N_{}$~ & ~$H_{SM}$~  & ~$H_1^*$~& ~$H_2$~
  \\\hline 
 $SU(2)_L$ & $\bm{2}$     & $\bm{1}$  & $\bm{1}$ & $\bm{2}$ & $\bm{2}$& $\bm{2}$    \\\hline 
$U(1)_Y$ & $\frac12$  & $-1$ & $0$  & $\frac12$  & $\frac12$ & $\frac12$      \\\hline
 $A_4$ & ${(1,1',1'')}$ & ${(1,1'',1')}$ & $3$ & $1$ & $1$ & $1$ \\\hline
 $-k$ & $0$ & $0$ & $-1$ & $0$ & $-1$ & $-5$   \\\hline
 $Z_3$ & $1$ & $1$ & $\omega$ & $1$ & $\omega^2$ & $\omega^2$   \\\hline
\end{tabular}
\caption{Fermionic and bosonic field content of the model and their charge assignments under $SU(2)_L\times U(1)_Y\times A_4$, where $-k$ is the number of modular weight. 
The quark sector is the same as that in the SM.}
\label{tab:fields}
\end{table}
\end{center}

\begin{center} 
\begin{table}[tb]
\begin{tabular}{|c||c|c|c|c|c|c|}\hline
 &\multicolumn{3}{c|}{Couplings}  \\\hline\hline
  & ~$Y^{(4)}_{\bf1}$~& ~$Y^{(2)}_{\bf3}$~ & ~$Y^{(6)}_{\bf3}$    \\\hline 
{ $A_4$} & ${\bf1}$ & ${\bf3}$ & ${\bf3}$      \\\hline
 $-k$ & $4$ & $2$ & $6$     \\\hline
\end{tabular}
\caption{Modular weight assignments for Yukawa couplings.}
\label{tab:couplings}
\end{table}
\end{center}

The modular forms of weight 2, {$(y_{1},y_{2},y_{3})$}, which transform
as a triplet of $A_4$, are written in terms of the Dedekind
eta-function, $\eta(\tau)$, and its derivative, $\eta'(\tau)$,
as~\cite{Feruglio:2017spp}
\begin{eqnarray} 
\label{eq:Y-A4}
y_{1}(\tau) &=& \frac{i}{2\pi}\left( \frac{\eta'(\tau/3)}{\eta(\tau/3)}  +\frac{\eta'((\tau +1)/3)}{\eta((\tau+1)/3)}  
+\frac{\eta'((\tau +2)/3)}{\eta((\tau+2)/3)} - \frac{27\eta'(3\tau)}{\eta(3\tau)}  \right), \nonumber \\
y_{2}(\tau) &=& \frac{-i}{\pi}\left( \frac{\eta'(\tau/3)}{\eta(\tau/3)}  +\omega^2\frac{\eta'((\tau +1)/3)}{\eta((\tau+1)/3)}  
+\omega \frac{\eta'((\tau +2)/3)}{\eta((\tau+2)/3)}  \right) , \label{eq:Yi} \\ 
y_{3}(\tau) &=& \frac{-i}{\pi}\left( \frac{\eta'(\tau/3)}{\eta(\tau/3)}  +\omega\frac{\eta'((\tau +1)/3)}{\eta((\tau+1)/3)}  
+\omega^2 \frac{\eta'((\tau +2)/3)}{\eta((\tau+2)/3)}  \right)\,.
\nonumber\label{eq:3rep}
\end{eqnarray}
Then, any couplings of higher weights are constructed from the
multiplication rules of $A_4$. One finds the following expressions:
\begin{align}
&Y^{(4)}_{\bf1}=y^2_1+2y_2y_3,\quad
Y^{(6)}_{\bf3}=
\left[\begin{array}{c}
y^3_1+2y_1y_2y_3 \\ 
y^2_1y_2+2y_2^2y_3 \\ 
y^2_1y_3+2y_3^2y_2 \\ 
\end{array}\right],\quad
Y'^{(6)}_{\bf3}=
\left[\begin{array}{c}
y^3_3 + 2y_1y_2y_3 \\ 
y^2_3y_1+2y_1^2y_2 \\ 
y^2_3y_2+2y_2^2y_1 \\ 
\end{array}\right].
\end{align}

To construct a nonzero neutrino mass matrix, we also need a quartic term
in the Higgs potential, {\it i.e.} $(H_{SM}^\dag H_1)(H_{SM}^\dag H_2)$,
which can be realized as follows:
\begin{align}
a_0\frac{Y^{(4)}_{\bf1}}{i(\tau^*-\tau)}(H_{SM}^\dag H_1)(H_{SM}^\dag H_2)+h.c.
\equiv \lambda_0(H_{SM}^\dag H_1)(H_{SM}^\dag H_2)+h.c.,\label{eq:kaelar}
\end{align}
where $a_0$ is an arbitrary complex number, {and the factor $\frac{1}{i(\tau^*-\tau)}$ is also important to keep this term invariant.}  Although it mixes the neutral
complex Higgs bosons, we assume that the mixing angle is very small
and the mass eigenstates are almost the flavor eigenstates, which we
denote as $\eta_{1,2}$.

\subsection{Neutrino mass matrix}
{Since $N$ transforms as a triplet under the $A_4$ group, its Dirac mass is invariant only in the case of $M_D(\bar N_{L_1}N_{R_1}+\bar N_{L_2}N_{R_2}+\bar N_{L_3}N_{R_3})$. Notice here the mass parameter $M_D$ includes the factor of $\frac{1}{i(\tau^*-\tau)}$, which is the same as the case in Eq.(\ref{eq:kaelar}). 
As a result,}
 the heavy Dirac neutrino mass matrix is
diagonal with the eigenvalue, $M_D$:
\begin{align}
{\cal M}&= {M_D}
\left[\begin{array}{ccc}
1 &0 & 0 \\ 
0 & 1 &0 \\ 
0& 0 & 1 \\ 
\end{array}\right] . \label{eq:diracm}
\end{align}

We write down the relevant interactions for the generation of the
neutrino mass matrix as
\begin{align}
-{\cal L}_\nu&=\bar\nu_L y_{N_R} N_R \eta_1^* 
+\bar N_Ly_{N_L} \nu^C_L \eta_2^*+\frac{\lambda_0v_H^2}{2}\eta_1\eta_2+h.c.,
\end{align}
where  
\begin{align}
y_{N_R} &=
\left[\begin{array}{ccc}
\alpha_\nu &0 & 0 \\ 
0 & \beta_\nu  &0   \\ 
0& 0 & \gamma_\nu  \\ 
\end{array}\right]
\left[\begin{array}{ccc}
y_{1} & y_{3} &y_{2} \\ 
y_{3} & y_{2} &y_{1} \\ 
y_{2} & y_{1} & y_{3} \\ 
\end{array}\right],\\
 y_{N_L} &=
\left[\begin{array}{ccc}
y_{6,1} & y_{6,3} &y_{6,2} \\ 
y_{6,3} & y_{6,2} &y_{6,1} \\ 
y_{6,2} & y_{6,1} & y_{6,3} \\ 
\end{array}\right]^*
\left[\begin{array}{ccc}
a_\nu &0 & 0 \\ 
0 & 0 & c_\nu   \\ 
0& b_\nu & 0  \\ 
\end{array}\right]
+
\left[\begin{array}{ccc}
y'_{6,1} & y'_{6,3} &y'_{6,2} \\ 
y'_{6,3} & y'_{6,2} &y'_{6,1} \\ 
y'_{6,2} & y'_{6,1} & y'_{6,3} \\ 
\end{array}\right]^*
\left[\begin{array}{ccc}
a'_\nu &0 & 0 \\ 
0 & 0  & c'_\nu   \\ 
0& b'_\nu & 0  \\ 
\end{array}\right],
\label{eq:mn}
\end{align}
where ${\bf Y^{(6)}_3}\equiv[ y_{6,1},y_{6,2},y_{6,3}]^T$, ${\bf
Y'^{(6)}_3}\equiv[ y'_{6,1},y'_{6,2},y'_{6,3}]^T$, and we impose the
perturbativity limit, ${\rm Max}[y_{N_{R,L}}]\lesssim\sqrt{4\pi}$, in
the numerical analysis.

Then, the neutrino mass matrix is calculated as
\begin{align}
&m_{\nu}=\kappa [y_{N_R}  y_{N_L} +(y_{N_R}  y_{N_L})^T]\equiv \kappa \tilde m_\nu,\\
&\kappa=\frac{\lambda_0^* v_{SM}^2}{32\pi^2M_D} F(M_D^2,\eta_1^2,\eta_2^2),\\
& F(a,b,c)=-a\frac{(b-c)a\ln a+(c-a)b\ln b+(a-b)c\ln c}{(a-b)(b-c)(c-a)},
\end{align}
where $v_{SM}\simeq$ 246 GeV is the VEV of the SM Higgs boson. Notice
that $\kappa$ does not depend on the flavor structure due to the modular
symmetry.

Thus, we determine $\kappa$ by
\begin{align}
(NH):\  \kappa^2= \frac{|\Delta m_{\rm atm}^2|}{\tilde D_{\nu_3}^2-\tilde D_{\nu_1}^2},
\quad
(IH):\  \kappa^2= \frac{|\Delta m_{\rm atm}^2|}{\tilde D_{\nu_2}^2-\tilde D_{\nu_3}^2},
 \end{align}
where $\tilde m_\nu$ is diagonalized by $V^\dag_\nu (\tilde m_\nu^\dag
\tilde m_\nu)V_\nu=(\tilde D_{\nu_1}^2,\tilde D_{\nu_2}^2,\tilde
D_{\nu_3}^2)$~\footnote{The diagonal elements of $\tilde D_{\nu}$ are dimensionless neutrino eigenvalues that is convenient to determine the neutrino mass ratios.} and $\Delta m_{\rm atm}^2$ is the atmospheric neutrino
mass-squared difference. Here, NH and IH stand for the normal hierarchy
and the inverted hierarchy, respectively.  Subsequently, the solar
neutrino mass-squared difference can be written in terms of $\kappa$ as
follows:
\begin{align}
\Delta m_{\rm sol}^2= {\kappa^2}({\tilde D_{\nu_2}^2-\tilde D_{\nu_1}^2}),
 \end{align}
which we compare with the observed value.
  
Working in the diagonal basis of the charged-leptons, one finds $U_{\rm
PMNS}=V_\nu$, which is parameterized by three mixing angles, $\theta_{ij}
(i,j=1,2,3; i < j)$, one CP violating Dirac phase, $\delta_{CP}$, and two
Majorana phases, $\{\alpha_{2}, \alpha_{3}\}$, as
\begin{equation}
U_{\rm PMNS} = 
\begin{pmatrix} c_{12} c_{13} & s_{12} c_{13} & s_{13} e^{-i \delta_{CP}} \\ 
-s_{12} c_{23} - c_{12} s_{23} s_{13} e^{i \delta_{CP}} & c_{12} c_{23} - s_{12} s_{23} s_{13} e^{i \delta_{CP}} & s_{23} c_{13} \\
s_{12} s_{23} - c_{12} c_{23} s_{13} e^{i \delta_{CP}} & -c_{12} s_{23} - s_{12} c_{23} s_{13} e^{i \delta_{CP}} & c_{23} c_{13} 
\end{pmatrix}
\begin{pmatrix} 1 & 0 & 0 \\ 0 & e^{i \frac{\alpha_{2}}{2}} & 0 \\ 0 & 0 & e^{i \frac{\alpha_{3}}{2}} \end{pmatrix},
\end{equation}
where $c_{ij}$ and $s_{ij}$ stand for $\cos \theta_{ij}$ and $\sin \theta_{ij}$, respectively. 
Then, the mixing angles are expressed in terms of the components of $U_{\rm PMNS}$ as
\begin{align}
\sin^2\theta_{13}=|(U_{\rm PMNS})_{13}|^2,\quad 
\sin^2\theta_{23}=\frac{|(U_{\rm PMNS})_{23}|^2}{1-|(U_{\rm PMNS})_{13}|^2},\quad 
\sin^2\theta_{12}=\frac{|(U_{\rm PMNS})_{12}|^2}{1-|(U_{\rm PMNS})_{13}|^2}.\label{eq:angles}
\end{align}
As for the CP phase and the Majorana phases, we have
\begin{align}
 \delta_{\rm CP}&=-{\rm arg}[\det U_{\rm PMNS}]+{\rm arg}[(U_{\rm PMNS})_{11}]+{\rm arg}[(U_{\rm PMNS})_{12}]\nonumber\\
 &\hspace{3ex}-{\rm arg}[(U_{\rm PMNS})_{13}]+{\rm arg}[(U_{\rm PMNS})_{23}]+{\rm arg}[(U_{\rm PMNS})_{33}],\\
 \frac{\alpha_2}{2}&={\rm arg}[(U_{\rm PMNS})_{12}]-{\rm arg}[(U_{\rm PMNS})_{11}],\\
 \frac{\alpha_3}{2}&=-{\rm arg}[\det U_{\rm PMNS}]+{\rm arg}[(U_{\rm PMNS})_{12}]+{\rm arg}[(U_{\rm PMNS})_{23}]+{\rm arg}[(U_{\rm PMNS})_{33}],\label{eq:phases}
\end{align}
where ${\rm arg}[*]$ denotes the argument of $*$.
{In our numerical analysis, we will implicitly use Eqs.~(\ref{eq:angles})-(\ref{eq:phases}) as output parameters to compare with experimental values below.}
We will adopt the neutrino experimental data at the 3$\sigma$ interval~\cite{Esteban:2018azc} as follows:
\begin{align}
&{\rm NH}: \Delta m^2_{\rm atm}=[2.431, 2.622]\times 10^{-3}\ {\rm eV}^2,\
\Delta m^2_{\rm sol}=[6.79, 8.01]\times 10^{-5}\ {\rm eV}^2,\label{eq:nuexp_NH}\\
&\sin^2\theta_{13}=[0.02044, 0.02437],\ 
\sin^2\theta_{23}=[0.428, 0.624],\ 
\sin^2\theta_{12}=[0.275, 0.350],\nn\\
&{\rm IH}: \Delta m^2_{\rm atm}=[2.413, 2.606]\times 10^{-3}\ {\rm eV}^2,\
\Delta m^2_{\rm sol}=[6.79, 8.01]\times 10^{-5}\ {\rm eV}^2,\label{eq:nuexp_IH}\\
&\sin^2\theta_{13}=[0.02067, 0.02461],\ 
\sin^2\theta_{23}=[0.433, 0.623],\ 
\sin^2\theta_{12}=[0.275, 0.350].\nn
\end{align}
Notice that the atmospheric mass-squared difference is
considered as an input parameter in our numerical analysis.

\subsection{Neutrinoless double beta decay}
The neutrinoless double beta decay may be observed by KamLAND-Zen in
future~\cite{KamLAND-Zen:2016pfg}.  The relevant effective mass can be calculated as
\begin{align}
\langle m_{ee}\rangle=\kappa|\tilde D_{\nu_1} \cos^2\theta_{12} \cos^2\theta_{13}+\tilde D_{\nu_2} \sin^2\theta_{12} \cos^2\theta_{13}e^{i\alpha_{2}}+\tilde D_{\nu_3} \sin^2\theta_{13}e^{i(\alpha_{3}-2\delta_{CP})}|.
\end{align}

\subsection{Lepton flavor violations}
LFVs are induced via $y_{N_{L,R}}$ and the most stringent constraint
arises from $\mu\to e\gamma$ process, which is bounded as ${\rm
BR}(\mu\to e\gamma)\lesssim 4.2\times 10^{-13}$~\cite{TheMEG:2016wtm,
Renga:2018fpd}.  
{The relevant Lagrangian is given by
\begin{align}
-{\cal L}_\nu&=\bar\ell_L y_{N_R} N_R \eta_1^- 
+\bar N_Ly_{N_L} \ell^C_L \eta_2^- +h.c.,
\end{align}
where the flavor indices are omitted.
}
Meanwhile, our theoretical formula is given by
\begin{align}
{\rm BR}(\mu\to e\gamma)&= \frac{3\alpha_{\rm em}}{16\pi{\rm G_F^2}}
\left| G(M_D,m_{\eta_1})\left( \sum_{a=1}^3 y_{N_{R_{1a}}} y^\dag_{N_{R_{a2}}} \right)
+ 
G(M_D,m_{\eta_2}) \left( \sum_{a=1}^3 y_{N_{L_{a1}}} y^\dag_{N_{L_{2a}}}\right)
  \right|^2,\\
 G(m_a,m_b)&\approx
 \frac{2 m_a^6+3 m_a^4 m_b^2-6 m_a^2 m_b^4+m_b^6 +12 m_a^4 m_b^2 \ln\left[\frac{m_b}{m_a}\right]} {12 (m_a^2-m_b^2)^4},
\end{align}
where ${\rm G_F}\approx 1.17\times10^{-5}$ GeV$^{-2}$ is the Fermi
constant and $\alpha_{\rm em}\approx1/129$ is the fine structure
constant. When the masses of $m_{\eta_{1,2}}, M_D$ are of the order of
100 GeV, the constraints for Yukawa couplings are found as
$y_{N_R}\approx y_{N_L}\lesssim{\cal O}(0.01)$~\footnote{In our numerical
analysis, we also consider the other possible processes such
$\tau\to e\gamma$ and $\tau\to \mu\gamma$. These experimental results
give the constraints for Yukawa couplings $\lesssim{\cal O}$(0.1).}.

{
In a similar way as in the above LFVs, we can also formulate the muon $g-2$ and electron $g-2$ as follows:
\begin{align}
\Delta a_\ell \sim -\frac{m_\ell^2}{(4\pi)^2} \left[G(M_D,m_{\eta_1})\left( \sum_{a=1}^3 y_{N_{R_{\ell a}}} y^\dag_{N_{R_{a\ell}}} \right)
+ 
G(M_D,m_{\eta_2}) \left( \sum_{a=1}^3 y_{N_{L_{a\ell}}} y^\dag_{N_{L_{\ell a}}}\right)\right],\label{eq:g-2}
\end{align}
where $\ell=e,\mu$.  In our numerical analysis, we consider the
constraints from the LFVs and the $g-2$. 
{If $m_{\eta_1}=m_{\eta_2}=M_D=$100 GeV, Eq.(\ref{eq:g-2}) is simplified as follows:
\begin{align}
|\Delta a_e|=6.86\times 10^{-15}\times \sum_{a=1}^3  \left( y_{N_{R_{1 a}}} y^\dag_{N_{R_{a1}}} + y_{N_{L_{a1}}} y^\dag_{N_{L_{1 a}}}\right)\lesssim 8.8\times 10^{-13},\\
|\Delta a_\mu|=2.95\times 10^{-10}\times  \sum_{a=1}^3  \left( y_{N_{R_{2 a}}} y^\dag_{N_{R_{a2}}} + y_{N_{L_{a2}}} y^\dag_{N_{L_{2 a}}}\right)\lesssim 2.61\times10^{-9},
 \end{align}
where the constraints are set so that the new contributions do not exceed the current discrepancy between the experimental values and the SM predictions.
These constraints can be expressed in terms of the Yukawa upper bounds as $y_{N_R}\approx y_{N_L}\lesssim 4.61$ for electron g-2 and 
 $y_{N_R}\approx y_{N_L}\lesssim 1.21$ for muon g-2, both of which are weaker than the constraint of $\mu\to e\gamma$.}
We will also show concrete values as a benchmark point.

It would be worthwhile to mentioning the other LFVs such as $\ell_i\to \ell_j\ell_k\bar\ell_\ell$,
and muonium-antimuonium oscillation.
In case of three body decay type of LFVs, we have two processes; penguin type and box one at one-loop level.
The contribution to penguin type is always smaller than the case of two body decay type of LFVs $\ell_i\to \ell_j\gamma$ discussed above. This is because this process is written in terms of  the formula of two body decay type of LFVs times some suppressed factor such as fine structure constant.  Thus we do not need to consider this process.
The contribution to the box-type of LFVs arises from the Yukawa couplings of $y_{N_{R,L}}$ would give some constraints in our parameter space. The stringent bound comes from $\mu\to ee\bar e$, and its branching ratio is less than $10^{-12}$. From this bound, we find the following relation~\cite{Toma:2013zsa}
\begin{align}
\sum_{a,b=1}^3\left[(y_{N_R})_{1a} (y^\dag_{N_R})_{a2} (y_{N_R})_{1b} (y^\dag_{N_R})_{b1}+
(y_{N_L})^\dag_{1a} (y_{N_R})_{a2} (y_{N_R})^\dag_{1b} (y_{N_R})_{b1}\right]
\lesssim 
\left(\frac{M_D}{5.83{\rm TeV}}\right)^2,
\end{align}
It suggests that $\sum_{a,b=1}^3\left[(y_{N_R})_{1a} (y^\dag_{N_R})_{a2} (y_{N_R})_{1b} (y^\dag_{N_R})_{b1}+
(y_{N_L})^\dag_{1a} (y_{N_R})_{a2} (y_{N_R})^\dag_{1b} (y_{N_R})_{b1}\right]\lesssim 8.82\times10^{-4}$,
if we fix $M_D=100$ GeV. Therefore, this bound is weaker than the one of $\mu\to e\gamma$ and we do not need to take it into account.
%
Muonium-antimuonium oscillations are induced via $y_{N_R}$ and $y_{N_L}$ via box diagram.
In our estimation, we find the following bound~\cite{Conlin:2020veq}
\begin{align}
\sum_{a,b=1}^3(y^\dag_{N_L})_{2a} (y^\dag_{N_R})_{a2} (y_{N_R})_{1b} (y_{N_L})_{b1}
\lesssim
(4\pi)^2\times\left(\frac{M_D}{5.4{\rm TeV}}\right)^2.
\end{align}
It suggests that $\sum_{a,b=1}^3(y^\dag_{N_L})_{2a} (y^\dag_{N_R})_{a2} (y_{N_R})_{1b} (y_{N_L})_{b1}\lesssim 0.054$
if we fix $M_D=100$ GeV. Therefore, this bound is also very weak and we do not need to take it into account.

\subsection{DM analysis}
Here, we discuss DM, which we assume to be multi-component consisting of
$N_1,N_2$ and $N_3$.  Since our DM does not couple to nucleon at tree
level, we expect any bounds from direct detection experiments are not so
stringent~\footnote{We have interactions with nucleon via one-loop
diagrams, but we find our DM is totally safe for these processes. See
ref.~\cite{Schmidt:2012yg} in details.}. Thus, we just focus on the
relic density of our DM candidate.  Because of the nature of modular
$A_4$ symmetry in Eq.~(\ref{eq:diracm}), we have to consider the
co-annihilation processes among three degenerated $Ns$ in addition to
the annihilation. Supposing $n_\chi\equiv n_{N_1}= n_{N_2}= n_{N_3}$,
we find the same thermally averaged cross section formula appeared on
Appendix in ref.~\cite{Okada:2020oxh} except that the DM index, $i=1$,
of Yukawas $y_{N_{R,L}}$ is replaced by the sum over $i=1-3$.  We
constrain the relic density of DM to be $0.1193\pm0.0018$, which
corresponds to 2$\sigma$ confidential level of the Planck
result~\cite{Ade:2015xua}.

}

\section{Numerical analysis}\label{sec:num_analysis}
Here, we search for the allowed regions that satisfy the constraints on
the neutrino oscillation data given in Eqs.~(\ref{eq:nuexp_NH}) and
(\ref{eq:nuexp_IH}) and those on the {two body decay type
of }LFVs.  First, we set the ranges of input parameters as follows:
\begin{align}
&\tau=[-0.5+0.1i, 0.5+3i],\quad [\alpha_\nu,\beta_\nu,\gamma_\nu,a_\nu^{(')},b_\nu^{(')},c_\nu^{(')}]=[-1,1],\\
&[M_D,m_{\eta_1},m_{\eta_2}]=[1,10]\ {\rm TeV},\\
&{\rm NH}: \Delta m^2_{\rm atm}=[2.431, 2.622]\times 10^{-3}\ {\rm eV}^2,\
{\rm IH}: \Delta m^2_{\rm atm}=[2.413, 2.606]\times 10^{-3}\ {\rm eV}^2.
\end{align}
Here, the real part of $\tau$ has a periodicity of $1$ and the 3$\sigma$
interval of $\Delta m^2_{\rm atm}$ is used for the range of the scan.
%

\begin{figure}[tb]\begin{center}
\includegraphics[width=80mm]{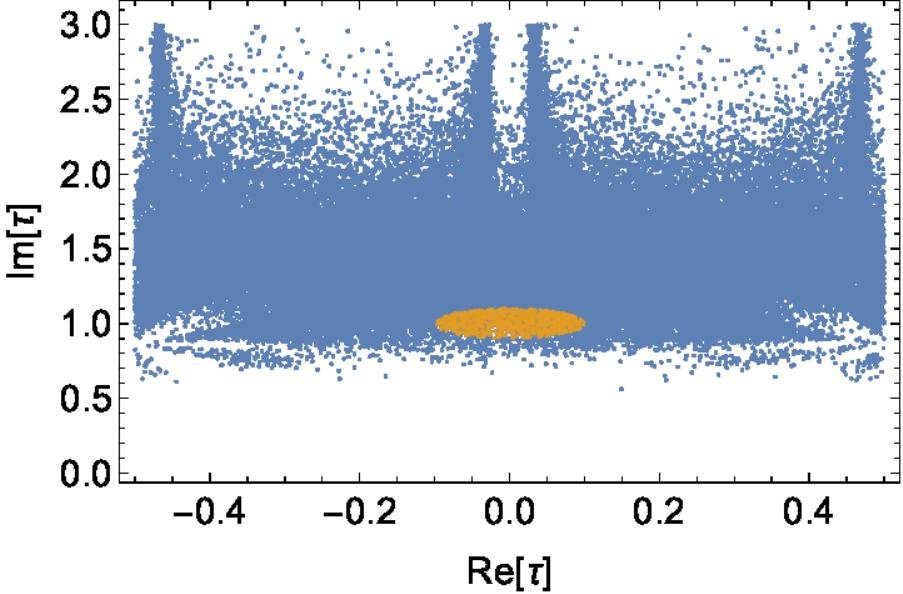}
\includegraphics[width=80mm]{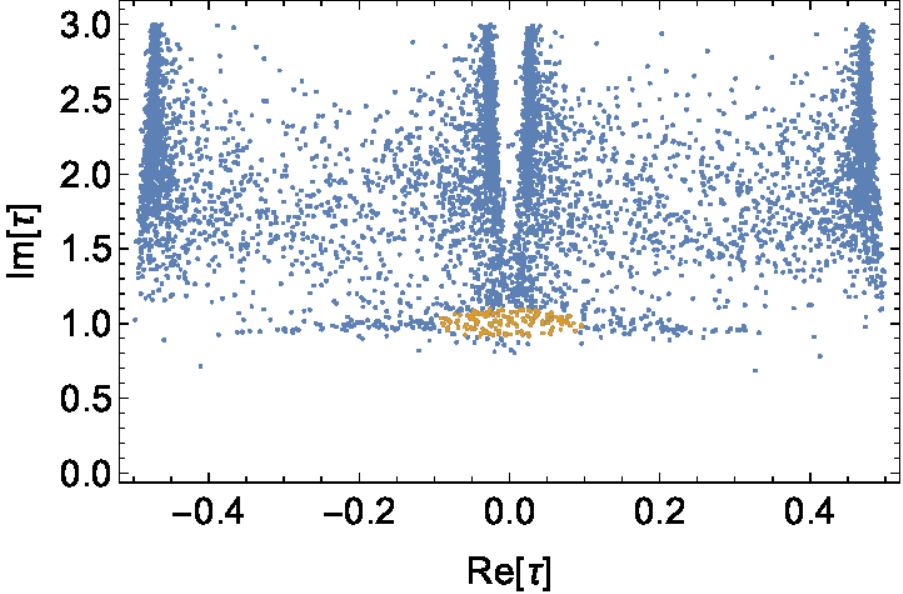} \caption{Allowed region of
$\tau$. The left panel is for the NH case and the right one is for the
IH case. The yellow points are the allowed points around
the fixed point.}  \label{fig:1}\end{center}\end{figure}

Fig.~\ref{fig:1} shows the allowed region of $\tau$. The left panel is
for the NH case and the right one is for the IH case. {All
the scattered points are allowed and the yellow points also satisfy
$|\tau-i|<0.1$, which corresponds to the fixed point that is favored by
string theory \cite{Kobayashi:2020uaj}. In the following, we use the
same colors for all figures.}
In both of NH and IH cases, the allowed regions have the similar shape,
{\it i.e.}  having four peaks at around ${\rm
Re}[\tau]=\pm0.1,\pm0.5$. A smaller ${\rm Im}[\tau]$ is preferred in the
NH case and a larger ${\rm Im}[\tau]$ is preferred in the IH case.
Here, we emphasize that the region around $\tau=i$ is allowed in both
cases and the same region is favored in a model of quark
sector~\cite{Okada:2019uoy}.  It suggests that our model may also
provide a viable and predictive quark model with the same value of
modulus $\tau$, although it is beyond our scope.

\begin{figure}[tb]\begin{center}
\includegraphics[width=80mm]{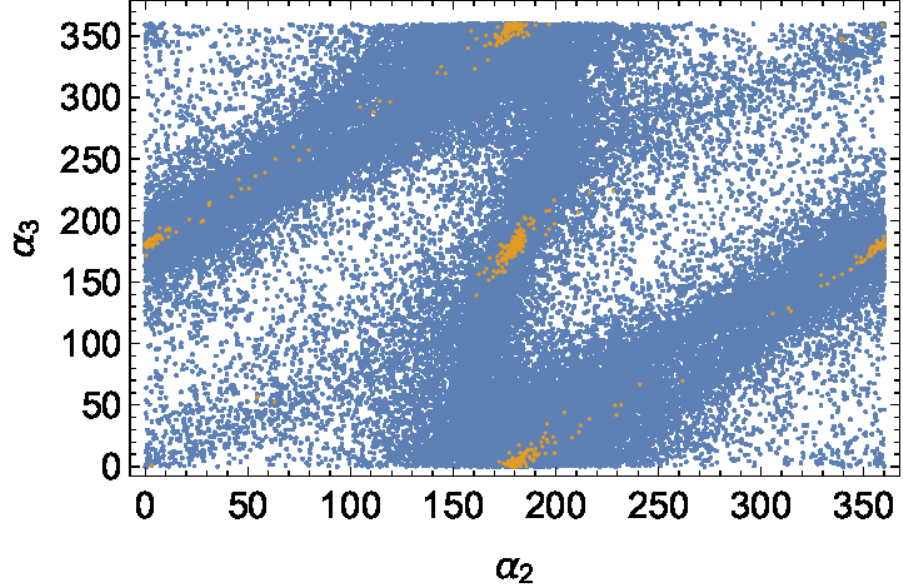}
\includegraphics[width=80mm]{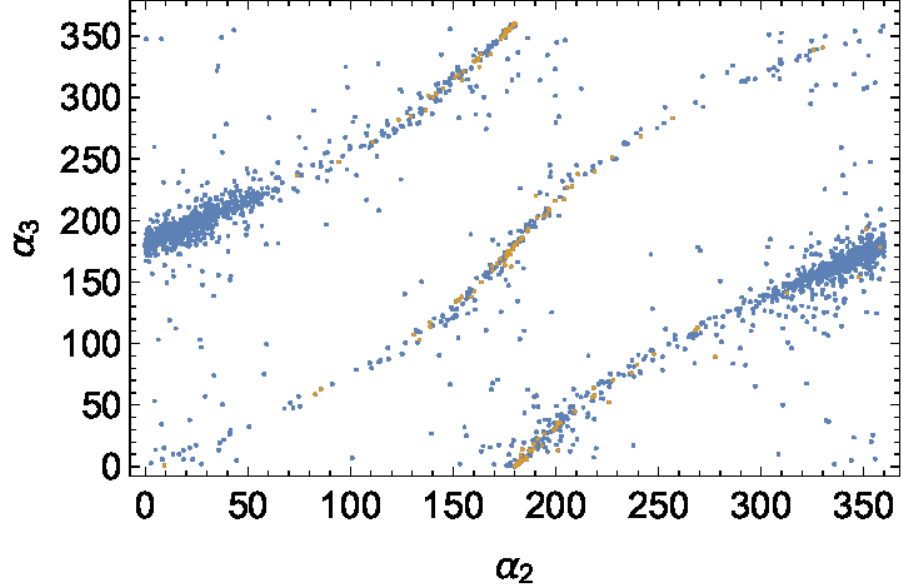}
\caption{Allowed region of $\alpha_2$ and $\alpha_3$. The left panel is for the NH case and the
right one is for the IH case. The yellow points are the allowed points around
the fixed point.}   
\label{fig:2}\end{center}\end{figure}

Fig.~\ref{fig:2} shows the allowed region of the Majorana phases,
$\alpha_2$ and $\alpha_3$. The left panel is for the NH case and the
right one is for the IH case.  We find characteristic denser regions in
both NH and IH.

\begin{figure}[tb]\begin{center}
\includegraphics[width=80mm]{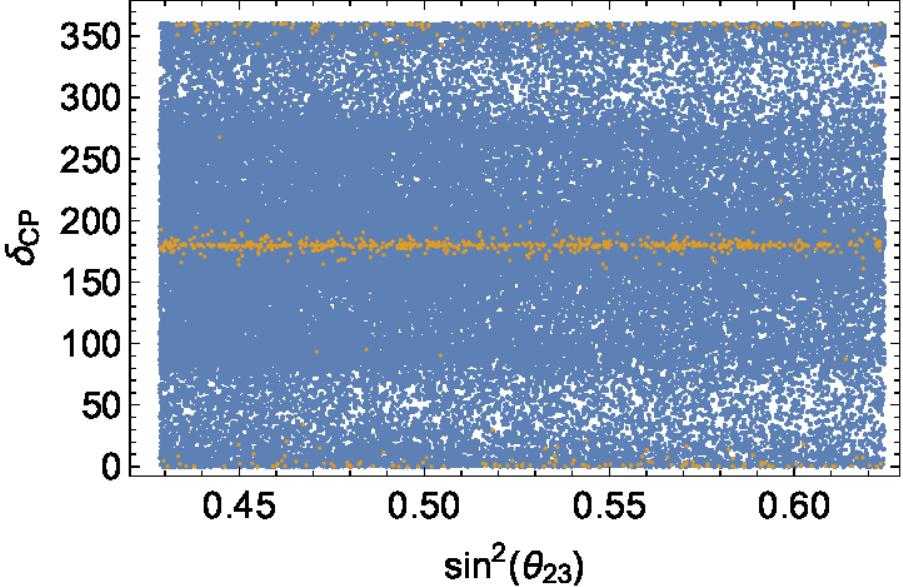}
\includegraphics[width=80mm]{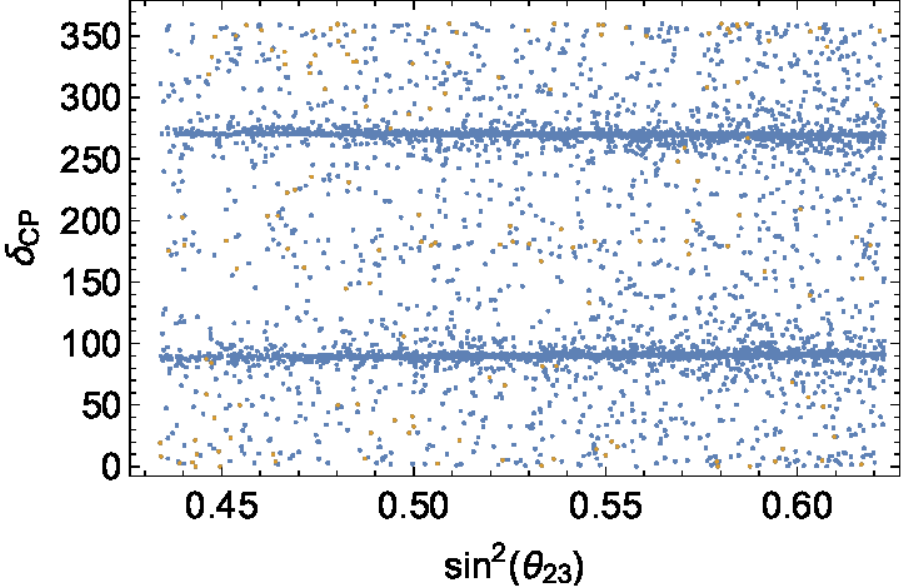}
\caption{Allowed region of $\sin^2\theta_{23}$ and $\delta_{\rm CP}$. The left panel is for the NH case and the
right one is for the IH case.  The yellow points are the allowed points around
the fixed point.}   
\label{fig:3}\end{center}\end{figure}

Fig.~\ref{fig:3} shows the allowed region of $\sin^2\theta_{23}$ and
$\delta_{\rm CP}$. The left panel is for the NH case and the right one
is for the IH case.  There are three denser regions at $\delta_{\rm
CP}=90^\circ,240^\circ, 270^\circ$ for NH, while two denser regions at
$\delta_{\rm CP}=90^\circ, 270^\circ$ for IH. Since the current best fit
value of the Dirac CP phase is around $270^\circ$, these predictions
seem to be consistent. Moreover, we find that a relatively smaller value
of $\sin^2\theta_{23}$ tends to be favored from the best fit value of
$\delta_{\rm CP}$ for the NH case and a relatively larger value for the
IH case.

\begin{figure}[tb]\begin{center}
\includegraphics[width=80mm]{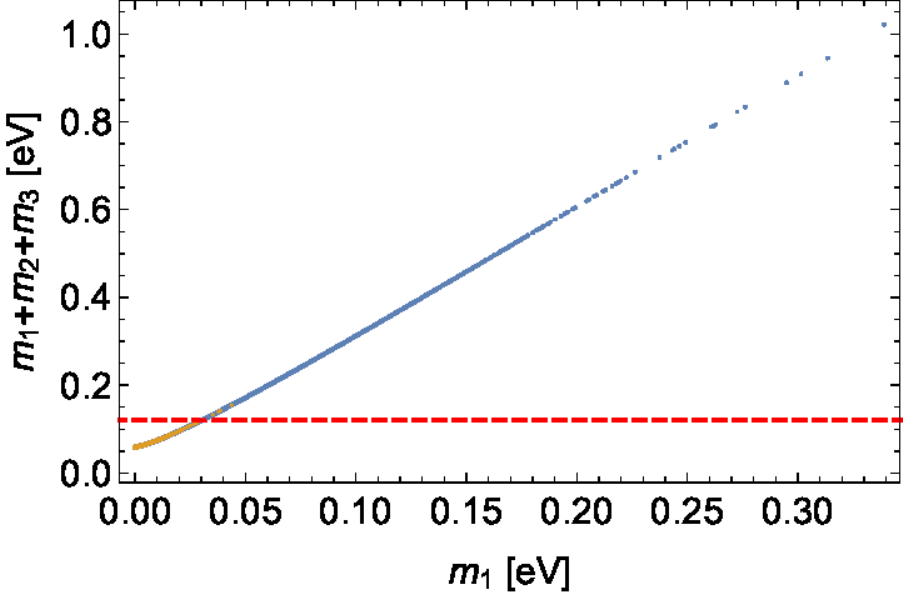}
\includegraphics[width=80mm]{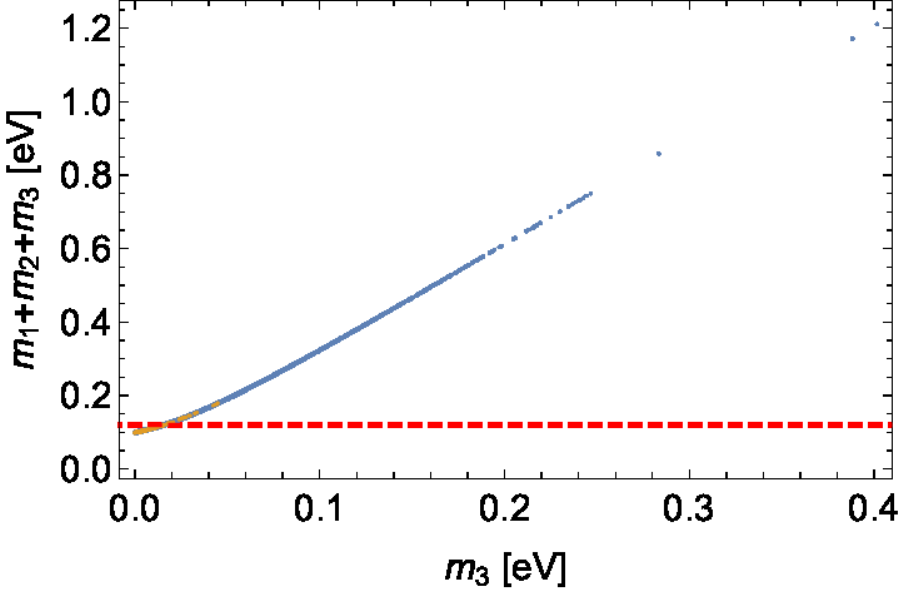}
\caption{Allowed region of the lightest neutrino mass and $m_1+m_2+m_3$. The left panel is for the NH case and the
right one is for the IH case.  The yellow points are the allowed points around
the fixed point.} 
\label{fig:4}\end{center}\end{figure}

Fig.~\ref{fig:4} shows the allowed region of the lightest neutrino mass,
$m_1$ for NH and $m_3$ for IH, and the sum of neutrino masses,
$m_1+m_2+m_3$. The left panel is for the NH case and the right one is
for the IH case. The horizontal red dotted line represents the
cosmological constraint on the sum of the neutrino masses, where $\sum_i
m_i\lesssim0.12$ eV~\cite{Aghanim:2018eyx, Vagnozzi:2017ovm}.  If
this constraint is considered seriously, we get a stronger constraint on
the lightest neutrino mass around $0.03$ eV for both NH and IH.

\begin{figure}[tb]\begin{center}
\includegraphics[width=80mm]{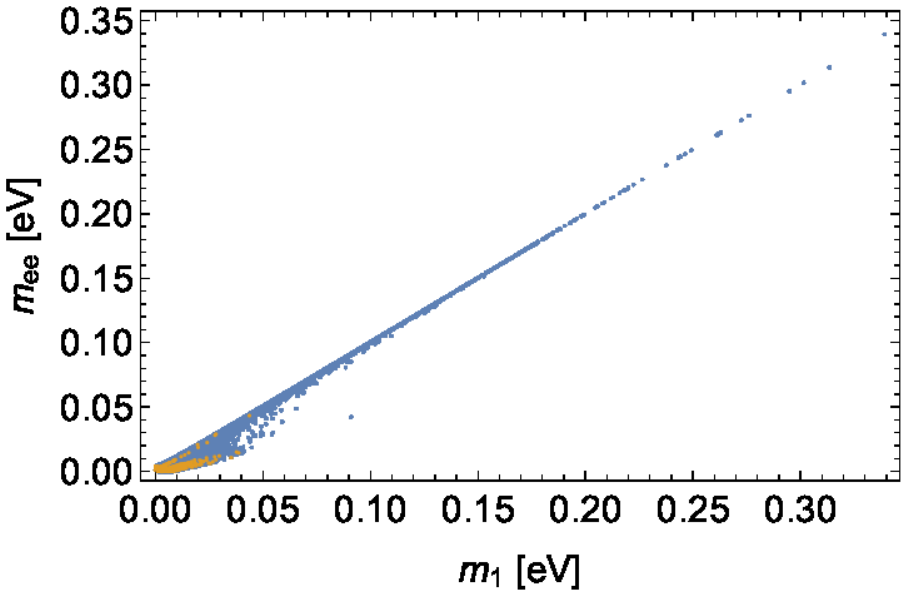}
\includegraphics[width=80mm]{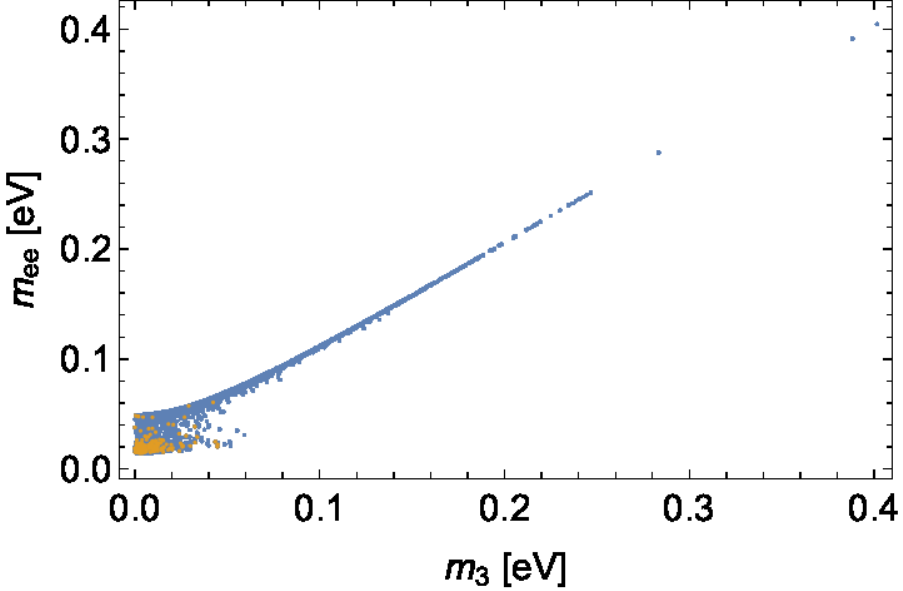}
\caption{Allowed region of $m_1$ and $m_{ee}$. The left panel is for the NH case and the
right one is for the IH case.  The yellow points are the allowed points around
the fixed point.}   
\label{fig:5}\end{center}\end{figure}

Fig.~\ref{fig:5} shows the allowed region of the lightest neutrino mass
and the effective mass for the neutrinoless double beta decay,
$m_{ee}$. The left panel is for the NH case and the right one is for the
IH case.  Considering the cosmological constraint shown in
Fig.~\ref{fig:4}, we obtain an upper bound on $m_{ee}$;
$m_{ee}\lesssim 0.04$ eV for NH and $m_{ee}\lesssim 0.08$ eV for IH.

\begin{figure}[tb]\begin{center}
\includegraphics[width=80mm]{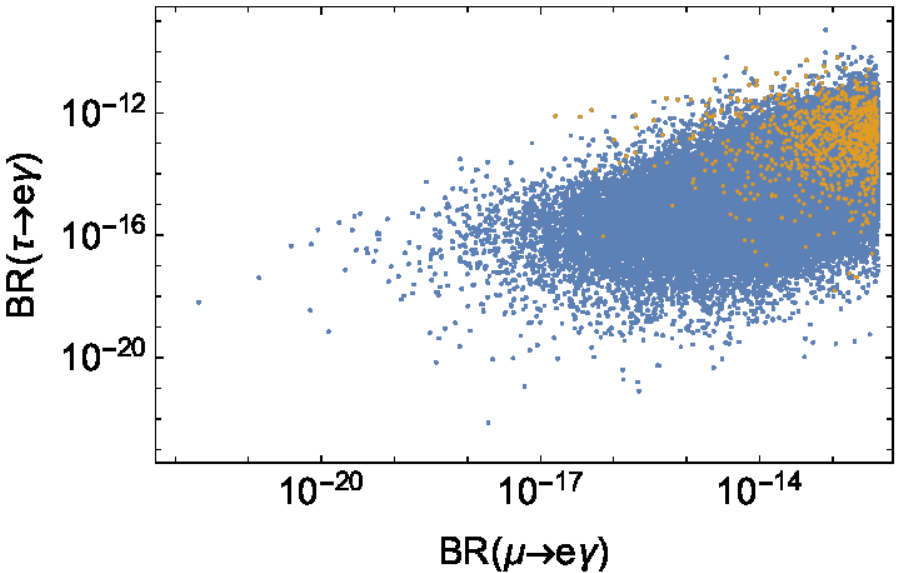}
\includegraphics[width=80mm]{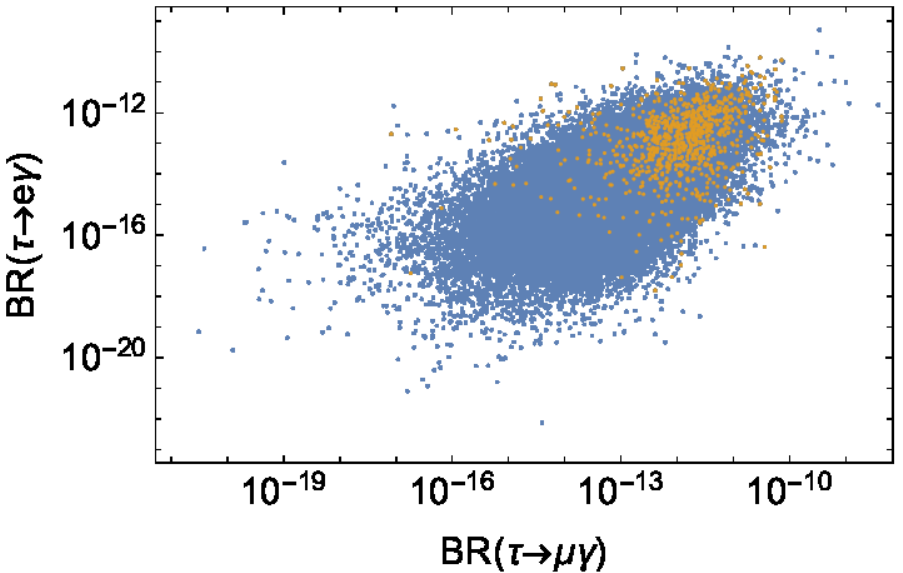}
\includegraphics[width=80mm]{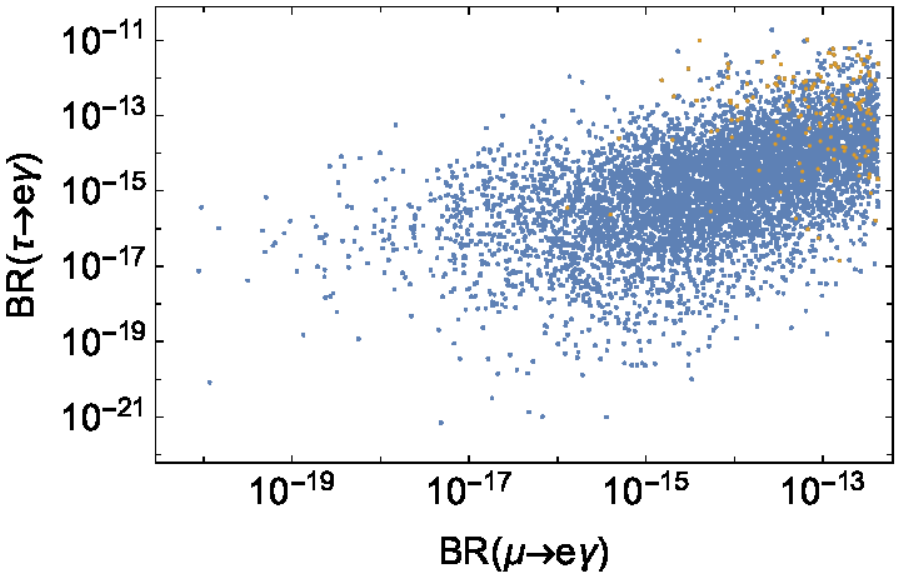}
\includegraphics[width=80mm]{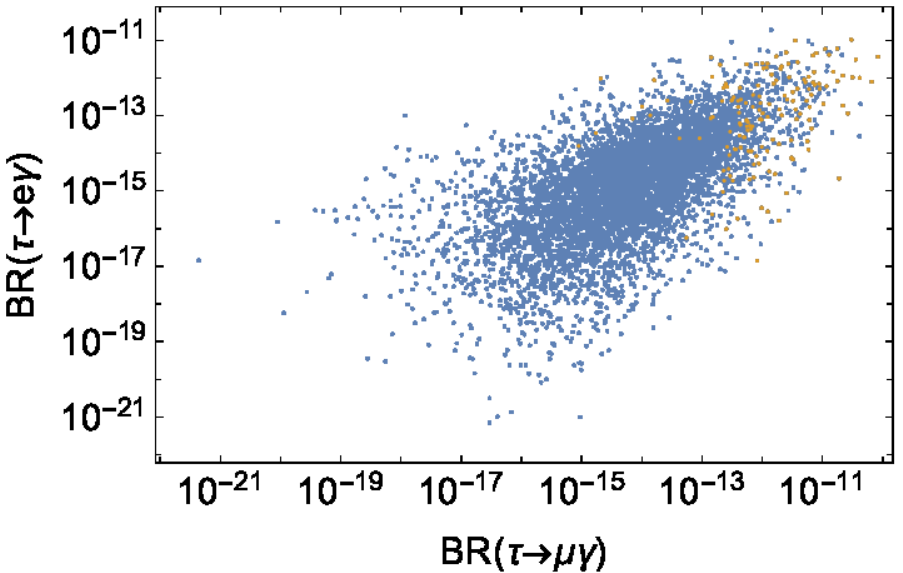}
\caption{The allowed points in the $\mu\to e\gamma$ vs $\tau\to e\gamma$ plane and in the $\tau\to\mu\gamma$ vs $\tau\to e\gamma$ plane. The top panels are for the NH case and the bottom ones are for the IH case. The yellow points are the allowed points around
the fixed point.}   
\label{fig:lfv}\end{center}\end{figure}

{In Fig.~\ref{fig:lfv}, we show the scattered plot for the
LFVs.  After the constraint on ${\rm BR}(\mu\to e\gamma)$ is imposed,
the typical size of ${\rm BR}(\tau\to e\gamma)$ and ${\rm BR}(\tau\to
\mu\gamma)$ are one or more orders of magnitude smaller than the current
constraints of $3.3\times10^{-8}$ and $4.4\times10^{-8}$,
respectively. For all the allowed points, the contribution to the muon
$g-2$ is $0<-\Delta a_\mu\lesssim10^{-11}$ and the contribution to the
electron $g-2$ is $0<-\Delta a_e\lesssim10^{-16}$. Since they are much
smaller than the observed anomalies, we will not further consider them.}

All the other oscillation parameters are found over all the 3$\sigma$
intervals of~Eq.(\ref{eq:nuexp_NH}) without structures. 

{Although it is difficult to explain all the structures
appearing in Fig.~\ref{fig:2} and \ref{fig:3}, we can interpret the
location of the yellow points.  In the case of $\tau=i$, $(y_1,y_2,y_3)$
in Eq.~(\ref{eq:3rep}) is simply given by
$Y_0(1,1-\sqrt{3},-2+\sqrt{3})$, where $Y_0\approx1.0225$, and hence the
neutrino mass matrix is real. Thus, as we can see from Fig.~\ref{fig:2}
and \ref{fig:3}, the yellow points are located around where the Dirac CP
phase and the Majorana phases become $0$ or $\pi$. If we go far from the
fixed point, the Yukawa couplings depend non-trivially on $\tau$ and the
shape of the blue points are determined mostly by the constraints on the
neutrino mixing angles.

When we include the analysis of our DM candidate, our parameter space
drastically reduces, while the prediction on lepton sector would not
change.  Thus, we just show a benchmark point to satisfy all the
experimental date at the fixed point as follows:
\begin{align}
{\rm Moduli}\colon&\tau\sim 0.0233+0.991 i,\nn\\
{\rm Yukawa}\colon&\alpha_\nu\sim0.0240002,\ \beta_\nu\sim 0.894747,\ \gamma_\nu\sim -0.382786, \nn\\
& a_\nu \sim0.042218,\
 b_\nu-\sim 0.784245,\
c_\nu-\sim 0.213312,\nn\\
&a'_\nu \sim -0.274272, \
b'_\nu \sim 0.400292, \
c'_\nu \sim 0.719127,\nn\\ 
{\rm Mass}\colon& M_D \sim 6919.92 {\rm GeV},
m_{\eta_1} \sim 5117.94 {\rm GeV},\
m_{\eta_2} \sim 3790.11,\nn\\
{\rm Quatic}\colon&\lambda_0\sim 6.77552\times10^{-10},\nn\\
{\rm Neutrino}\colon& m_1 \sim 1.80269\times10^{-12},\
m_2\sim 8.64233\times10^{-12},\
m_3 \sim 4.99623\times10^{-11},\nn\\
& \sin^2\theta_{12}\sim 0.574245,\
\sin^2\theta_{23}\sim 0.740345,\
\sin^2\theta_{13} \sim 0.146588,\nn\\
&\delta_{CP}\sim 3.13183,\
\alpha_{2} \sim 3.14909,\
\beta_{2}  \sim6.27842,\nn\\
{\rm LFV}\colon& {\rm BR}(\mu\to e\gamma)\sim 1.1601\times10^{-13},\
{\rm BR}(\tau\to e\gamma)  \sim 1.51088\times10^{-12},\nn\\
&{\rm BR}(\tau\to \mu\gamma) \sim 1.66469\times10^{-14},\nn\\
{\rm g-2}\colon&\Delta a_e\sim -1.30281\times 10^{-18},\ \Delta a_\mu \sim -8.41567\times10^{-13},\nn\\
{\rm DM}\colon& 
 \Omega h^2 \sim 0.118351.
\end{align}

}

\section{Conclusion and discussion}\label{sec:conclusion}
We have extended the radiative seesaw model proposed in our previous
work~\cite{Okada:2020oxh} by applying a modular $A_4$ symmetry in order
to increase predictability in the lepton sector.  After the formulation
of the neutrino sector and LFVs, we have executed a numerical analysis
to find parameters that satisfy the constraints from the neutrino
oscillation data and LFVs. We have found several preferences in the
allowed parameter regions.  We highlight some prominent points in the
following.
 \begin{enumerate}
\item 
We find characteristic denser allowed regions of the Majorana phases.
\item
There are three denser regions at $\delta_{\rm
CP}=90^\circ,240^\circ, 270^\circ$ for NH, while two denser regions at $\delta_{\rm
CP}=90^\circ, 270^\circ$ for IH. Since the current best fit value of the Dirac
CP phase is around $270^\circ$, these predictions seem to be
consistent. Moreover, one might find that a relatively smaller value of $\sin^2\theta_{23}$
tends to be favored from the best fit value of $\delta_{\rm CP}$
for the NH case and a relatively larger value for the IH case.

\item
We obtain an upper bound on
$m_{ee}$; $m_{ee}\lesssim 0.04$ eV for NH and
$m_{ee}\lesssim 0.08$ eV for IH.
 \end{enumerate}
 
 {Especially, we focus on a fixed point at $\tau=i$, which
 is favored by string theory.  Around this point, we explored our
 allowed parameter region and we have found that Dirac CP phase and the
 Majorana phases favor the values of $0$ or $\pi$. This originates from
 a fact that $(y_1,y_2,y_3)$ becomes real at $\tau=i$.  Finally, we have
 shown a benchmark point including DM analysis.  }

\begin{acknowledgements}
This research was supported by an appointment to the JRG Program at the APCTP through the Science and Technology Promotion Fund and Lottery Fund of the Korean Government. This was also supported by the Korean Local Governments - Gyeongsangbuk-do Province and Pohang City (H.O.). 
H.O.~is sincerely grateful for the KIAS member. Y.S.~is supported by
Grant-in-Aid for Scientific research from the Ministry of Education,
Science, Sports, and Culture (MEXT), Japan, No.~16H06492.
\end{acknowledgements}

\end{document}